\documentclass[12pt]{iopart}
\usepackage{iopams}
\usepackage{graphicx}
\begin{document}

\newcommand{\be}{\begin{equation}}
\newcommand{\ee}{\end{equation}}

\title{Johnson-Nyquist noise and the Casimir force between real metals at nonzero temperature}

\author{Giuseppe Bimonte}

\address{Dipartimento di Scienze Fisiche Universit\`{a} di Napoli Federico
II\\ Complesso Universitario MSA Via Cintia I-80126 Napoli Italy;\\
INFN, Sezione di Napoli, Napoli, ITALY}

\ead{Bimonte@na.infn.it}

\begin{abstract}
It is well known since a long time that all lossy conductors at
finite temperature  display an electronic noise, the
Johnson-Nyquist noise,  arising from the thermal agitation of
electric charges inside the conductor. The existence of this noise
implies that two nearby discharged conductors at finite temperature
should repel each other, as a result of the electrodynamic interaction
between  the Johnson-Nyquist currents in either conductor and the
eddy currents they induce in the other.
It is suggested that this force is at the origin of the recently discovered
large repulsive
correction to the thermal Casimir force between two lossy metallic plates.
Further support for this physical picture is obtained by studying a simple system of
two linear noisy antennas.
Using elementary
concepts from circuit theory, we show that the repulsive force
engendered by the Johnson-Nyquist noise results  in the same kind of thermodynamic inconsistencies
found in the Casimir problem.
We
show that all
inconsistencies  are however resolved if account is taken of capacitive
effects associated with the end points of the antennas. Our findings
therefore suggest that capacitive effects resulting from the finite size of
the plates, may be essential for a resolution of the
analogous problems met in the thermal Casimir effect.
\end{abstract}


\section{Introduction}

In recent years, great advances in experimental techniques have
stimulated an intense theoretical and experimental activity on the
Casimir effect, and more in general on dispersion forces. For
recent reviews on this rapidly evolving field,  we address the
reader to Refs.\cite{bordag}. The   present contribution   reports
on our recent work \cite{bimonte} on a theoretical conundrum that
attracted much attention in the last few years, and is not yet
resolved as we write. The problem is that of determining the
influence of temperature on the Casimir force between two
conductors, when the latter are treated as real metals, i.e.
taking account of ohmic losses. This  is not an  academic
question, because the  increasing accuracy of Casimir force
measurements is rapidly approaching the point, where these thermal
corrections should become measurable. In fact, in a related
context, the effect of temperature on the Casimir-Polder
interaction between a Bose-Einstein condensate and a dielectric
surface has been recently observed \cite{obrec}. The problem of
the influence of temperature on the Casimir force   is an old one,
for it was already discussed in some detail,  both for dielectrics
and conductors, by Lifshitz in his seminal paper on the theory of
molecular attractive forces between solids \cite{lifs}. As its is
well known, in this macroscopic theory the bodies are
characterized by means of  a dielectric permittivity
$\epsilon(\omega)$ (we shall not consider in what follows magnetic
materials). When dealing with conductors, Lifshitz pointed out
that for plate separations $a$ of the order of one micron, the
important region of frequencies is $\omega/c \simeq 1/a$,
corresponding to the infrared part of the spectrum. At such high
frequencies ohmic losses are irrelevant, and accordingly Lifshitz
took for the dielectric function of the metal the expression
appropriate in the infrared, i.e. the plasma model
$\epsilon_P(\omega)$ \be
\epsilon_P(\omega)=1-{\Omega_p^2}/{\omega^2}\;,\label{plasma}\ee
where $\Omega_p$ is the plasma frequency.  Lifshitz used it to
estimate the leading temperature correction to the  Casimir
pressure. He observed in passing that for larger separations, i.e.
for lower frequencies, the plasma dielectric function should be
replaced by \be\epsilon= 4 \pi i \sigma\,/\omega\;,\label{low}\ee
where $\sigma$ is the ordinary electrical conductivity of the
metal, but no explicit computations were performed for this case.
After many years, the question  was
  revived  in Ref. \cite{sernelius}, which contains the
first explicit computation of the thermal correction to the
Casimir pressure, taking account of dissipation. For this purpose,
when evaluating Lifshitz formula  giving the Casimir pressure
between two {\it infinite} plates at finite temperature, the
authors of \cite{sernelius} used the well-known Drude model
\be\epsilon_D(\omega)=1-\frac{\Omega_p^2}{\omega \,(\omega + i
\gamma)}\;,\label{drude}\ee which provides a smooth interpolation
between Eq. (\ref{plasma}) and Eq. (\ref{low}) (the relaxation
frequency $\gamma$ in Eq. (\ref{drude}) is related the
conductivity as $\gamma=\Omega_p^2/(4 \pi \sigma)$). Replacement
of Eq. (\ref{plasma}) by Eq. (\ref{drude}) in Lifshitz formula
produced an unexpected result because,  in the case of good
conductors, instead of the expected small correction, Lifshitz
formula predicted a large {\it repulsive} correction, to the
extent that for large separations the Casimir force was halved
with respect to the predicted result for the model without
dissipation. In fact, the magnitude of this repulsive correction
is independent of the actual magnitude of the dissipation,
provided that some dissipation is present.
  The mathematical reason for this result is clear
\cite{sernelius}, because  Lifshitz formula for the Casimir
pressure at finite temperature consists of a sum over imaginary
discrete frequencies $\xi_n=2 \pi n\, k_B T/\hbar\;,\;n=0,1,\dots$
(the so-called Matsubara frequencies). Its first term, for  $n=0$,
provides the leading contribution at large temperature and/or
separations and since it involves the quantity $A=\lim_{\xi
\rightarrow 0} \epsilon(i \xi)\, \xi^2$,  the magnitude of this
term depends crucially on the behavior of the dielectric function
at zero frequency. Even for infinitesimal amounts $\gamma$ of
dissipation in Eq. (\ref{drude}), the quantity $A$ jumps
discontinuously from the plasma-model value $\Omega_p^2$ to zero,
resulting in a discontinuous change in the thermal correction to
the Casimir pressure, as soon as dissipation is turned on.

Another  problem was later pointed out in \cite{romero}, where it
was found that for metallic plates without impurities,
straightforward insertion of the Drude model into Lifshitz formula
results in a violation of the Nernst heat theorem, while no such
inconsistencies arise if the plasma model Eq. (\ref{plasma}) is
used. Analogous thermodynamical inconsistencies have also been
found in the case of dielectrics, when a Drude-like term is added
to the permittivity function to account for their small dc
electric conductivity \cite{most3}.
These puzzling results stimulated a lively debate among the
experts, and several different attitudes have emerged (for a
survey, see Refs.\cite{most2, milton} and the contribution to
these Proceedings by V. M. Mostepanenko). At the moment of this
writing, the experimental situation is not yet clear. Recent
measurements of the Casimir force in the submicron range
\cite{decca}, where thermal effects are small anyhow, appear to
exclude the repulsive thermal correction implied by the Drude
model and seem to be in agreement with the plasma model
prediction, but the actual level of precision achieved by these
experiments  has been questioned by some authors \cite{milton}.
Very likely, for a definitive experimental resolution of the
problem, we shall have to wait until the measurements will be
extended to separations of a few microns, for which rival
theoretical approaches give largely different predictions for the
thermal Casimir force.

\section{The physical origin of the large thermal correction}

In what follows, we shall   argue that the jump in the thermal
Casimir force predicted by Lifshitz formula, is not a mere
mathematical accident but it rather signals that, when dissipation
is turned on, a {\it new} physical phenomenon, specifically
connected with dissipation in conductors, is sneaking into the
problem of the thermal Casimir force.  In our view, discovering if
such a new phenomenon is actually present is of great importance,
because this might help us understanding if and when Lifshitz
theory can be used to describe it.

An important hint is obtained  from Refs. \cite{lamor}, where the
problem was investigated using an equivalent mathematical
formulation of Lifshitz theory, that allows a clean separation of
the contribution of thermal photons from that of zero-point
fluctuations. In this way, it was shown that dissipation has a
small influence on the contribution from zero-point fluctuations,
and that the large thermal repulsive correction  found in
\cite{sernelius} originates from thermally excited evanescent
photons, with transverse (TE) polarization. It is important to
note that no such waves exist when dissipation is absent, as it
occurs when the plasma model is used instead of the Drude model.
The important (real) frequencies of these fluctuations are of
order $\tilde{\omega}=\gamma (\omega_c\,/\Omega_p)^2$, where
$\omega_c=c/a$ is the characteristic frequency of the cavity
\cite{esquivel}. For typical Casimir experiments, and even more so
at low temperature, when $\gamma$ is small, $\tilde{\omega}$ is
much less than $\omega_c$ and therefore these near fields have
nothing to do with the cavity eigen-modes whose zero-point
energies are at the origin of the Casimir effect proper. In fact,
since typically $\tilde{\omega} \ll k_B \,T/\hbar$ the repulsive
force  associated with these TE evanescent fields is   a classical
effect \cite{esquivel}.

We can ask   what produces these fields.  At the low frequencies
involved retardation effects are negligible and the relevant TE
evanescent fields represent {\it quasi-static magnetic fields}. It
is natural to imagine that the source of these magnetic fields are
thermally excited electric currents  inside the plates, and one is
immediately led to identify them with the currents that are at the
origin of the familiar Johnson-Nyquist noise \cite{john}. As it is
well known, this is the electronic noise characteristic of {\it
resistive} conductors at finite temperature,  originating from the
thermal agitation of charges inside a conductor. A striking
feature of this noise is that it exists only in conductors that
display some electrical resistance, and it is absent in strictly
dissipationless conductors, explaining  why we do not see it if we
use for the metal a model with no dissipation, like the plasma
model in Eq. (\ref{plasma}). The repulsive correction to the
thermal Casimir force found in \cite{sernelius} appears now as the
natural effect of the electrodynamic interaction between the
Johnson currents in either plate and the eddy currents induced
in the other plate. From this new perspective, the whole question
appears basically as a slightly unconventional  problem in
electrical engineering, i.e. the problem of the mechanical
interaction between two nearby noisy (extended) antennas.
In order to put to a test these ideas, in the next Section
we shall investigate the mechanical force existing between  two
linear noisy antennas. Since two wires do not form a cavity, by
considering this simple system we get rid of the spurious
complication added by the eigen-modes of a Casimir apparatus, and
we can  isolate the features of the
interaction  caused by the Johnson-Nyquist noise.

\section{Mechanical interaction between two noisy linear antennas}

Let us consider a system of two identical discharged linear  antennas,
consisting of two pieces ${\cal C}_1$ and ${\cal C}_2$ of thin
metallic wire at temperature $T$, displaced by an amount $\vec{a}$
from each other (we consider the orientations of the wires as
fixed once and for all). We assume that the thickness of the wires is much smaller
than the separation $a$, and that their length $L$ is small
compared with the wavelength $\lambda$ of the radiated e.m.
fields, such that we can describe the two antennas  as two
concentrated resistors with resistance $R$ (depending solely on
the temperature). As it is well known \cite{john}, the Johnson-Nyquist
noise in a noisy resistor can   be described by replacing the
actual resistors by noiseless ones, connected each to a noise
generator producing a random e.m.f. ${\cal E}_i(t),\;i=1,2$ with
spectrum: \be \langle {\cal E}_i(\omega)\,{\cal
E}_j^*(\omega')\rangle=4 \pi k_B T\,R\,E(\omega/\omega_T)\,
\delta(\omega-\omega')\,\delta_{i\,j}\;,\label{spec}\ee where
angle brackets denote statistical averages, $i,j=1,2$,
$\omega_T=k_B T/\hbar$ and $E(y)=y\,(e^y-1)^{-1}$.

Obviously, besides a resistance, the two
antennas are characterized by a self-inductance ${\cal L}$, and a
mutual inductance  ${\cal M}(\vec{a})$ depending on the separation
$\vec a$. For the moment we shall not consider that the antennas, having a finite length, possess also
capacitances $C$ associated with their end-points. One might say that neglecting these edge-effects is correct,
as far as $a \ll L$, analogously to what
is always done in the Casimir problem, where the plates are considered to be infinite, as in Lifshitz theory,
because one assumes that for plate separations much smaller than the plates size,
edge effects are negligible. The system of
two antennas is then described by the following circuit equations:
\begin{eqnarray}
{\cal L} \,\frac{d i_1}{dt}+{\cal M}(\vec{a})\frac{d i_2}{dt}+R\,
i_1 &=& {\cal E}_1(t)\;, \nonumber\\
{\cal L}\, \frac{d i_2}{dt}+{\cal M}(\vec{a})\frac{d i_1}{dt}+R\,
i_2 &=& {\cal E}_2(t)\;.\label{sys}
\end{eqnarray}
From classical electrodynamics \cite{jacks}, one learns that the
force ${\vec F}(\vec{a})$ between two inductances   can be written
as: \be {\vec F} (\vec{a})=\langle i_1 \,i_2 \rangle\,
\vec{\nabla}_a {\cal M}(\vec{a})\;.\ee The correlator $\langle i_1
\,i_2 \rangle$ can be easily computed by taking the time-Fourier
transform of Eqs.(\ref{sys}) and using Eq. (\ref{spec}). After
some computations, we obtain for the force the simple formula: \be
{\vec F}= -k_B T\,H\,\vec{\nabla}_a (m^2)\;,\label{force}\ee where
$m= {\cal M}/{\cal L}$ and $H$ is the quantity: \be
H(\omega_R/\omega_T,\,m^2)=\frac{1}{\pi} \int_0^{\infty} d  x\, x
E\left(x\,\frac{ \omega_R}{\omega_T}\right)\,{\rm Im}\,\left[(1-i
x)^2+x^2 m^2\right]^{-1}\;,\label{H}\ee   where  $x=\omega/\omega_R$, with $\omega_R=R/{\cal
L}$ (It should be noted that the frequencies $\omega$ that
contribute to $H$ are in the range $0<\omega <
\min\{\omega_R,\omega_T\}$, and therefore the long wavelength
approximation used to write Eqs. (\ref{sys}) is justified,
provided either $\omega_R$ or $\omega_T$ are small compared to
$c/L$).

  It is easy to
verify that $H$ is positive definite, and therefore, since ${\cal
M}^2$ decreases as $a$ increases, the force is  {\it repulsive}.
Moreover, since   for small $R$'s $E$ in Eq. (\ref{H}) becomes
one,  and then $H$ becomes   independent of $R$,
it follows that in the limit of zero resistance the force attains the non-zero value \be
\lim_{R \rightarrow 0}{\vec F}= -k_B T\, H(0,m^2)\,\vec{\nabla}_a
(m^2) \, \;.\label{lowR}\ee However, it should be noted that the
force is zero if we consider strictly dissipationless resistors.
Consider now the
free energy ${\cal F}$ associated with the force $\vec F$. Since ${\vec F}=-\vec{\nabla}_a {\cal
F}$, we obtain from Eq.(\ref{force}) and Eq.(\ref{H}) \be {\cal
F}=\frac{k_B T}{\pi} \int_0^{\infty}\frac{d x}{x}\, E
\left(x\, \frac{\omega_R}{\omega_T}\right)\,{\rm Im}\log
\left[1+\left(\frac{x \,m}{1-i
\,x}\right)^2\right]\,.\label{freen} \ee  If the wires have no impurities, at
liquid Helium temperatures and below, the resistance $R(T)$
approaches zero as $T^2$ \cite{kittel}. Therefore for $T \rightarrow 0$, the ratio $\omega_R/\omega_T$ approaches zero,
the integrand in Eq. (\ref{freen}) becomes
independent of $T$ and we find that the free energy is of the form
\be {\cal F} \approx g(m^2)\,k_B T \; \label{freeuno}\ee where $g(z)$ is a
positive function (because the imaginary part of the argument of
the logarithm in Eq. (\ref{freen}) is  positive definite).
Recalling that the entropy $S$ is $S=-\partial {\cal F}/\partial
T$, we then find: \be\lim_{T \rightarrow 0}S=- k_B\, g(m^2) \;\equiv
S_0 <0.\label{entr}\ee Since $S_0$ depends on the separation among
the wires through $m^2$, this result represents a violation of
Nernst heat theorem.

Summarizing, we have found that the Johnson-Nyquist noise in the
antennas gives rise to a repulsive force, that does not vanish for
vanishing resistance, is zero for strictly  zero resistance and violates
the Nernst heat theorem in the case of resistors with no impurities.
These are exactly  the same features displayed by  the large repulsive thermal correction to the Casimir force
between lossy plates, described in the Introduction. This analogy strongly
supports the idea that the physical phenomenon behind
this    correction to the Casimir force is actually the Johnson noise in the plates.
\begin{figure}
\includegraphics{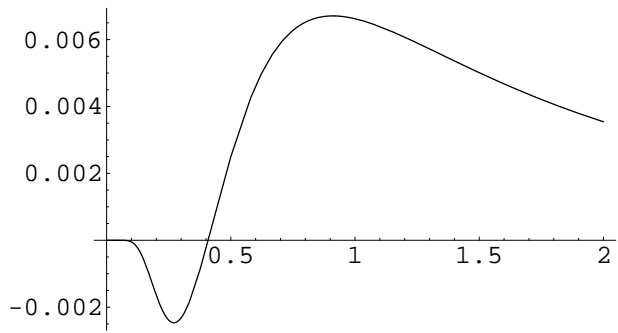}
\caption{\label{fig1}  Plot of the   free energy (in units of
$\hbar \omega_C$) as a function of $t=k_B T/(\hbar \omega_C)$. See
text for details.}
\end{figure}

\section{The role of the edges}

Given for granted that the Johnson-Nyquist noise exists and that it should give rise to a repulsion
between two nearby conductors, the question arises whether the puzzling features of the  force obtained in the
previous Section are the consequence of some incorrect assumption.
One of the assumptions that we made at the beginning of the previous Section
is in fact suspicious, when we said that for small separations it is correct to neglect the capacitances
associated with the end-points
of the antennas. It is worth then to repeat the computations, taking full account of these capacitances,
and see what happens.
This is easy to accomplish, for all we have to do is to include the capacitances of the antennas in the circuit equations, Eq.
(\ref{sys}).
The resulting equations can be easily solved and one finds that, {\it
independently of how large the
capacitances are}, the results obtained in the previous Section change drastically in some
important respects. Skipping all details, that
can be found in Ref. \cite{bimonte}, we just list the main points.
First, the force now vanishes in
the zero resistance limit, and thus one recovers smoothly the ideal metal
result. The effect of the capacitances on the magnitude of the force for finite values of the
resistance, depends of course on the value of the capacitance, and will not discussed here.
The most remarkable  change   is however   seen
at low temperature, for instead of Eq. (\ref{freeuno}) we obtain now  \be {\cal F}=- \frac{16
\pi^5\,m^2}{63}\,\left(\frac{k_B T}{\hbar \,\omega_C}\right)^6
\,\hbar\, \omega_R\;,\ee where $\omega_C=1/\sqrt{{\cal L}
\,C}$. Obviously, the entropy is now positive, and
vanishes as $T\rightarrow 0$, as required by Nernst theorem.
However, numerical computations show that the entropy is {\it
negative at intermediate} temperatures. This can be seen from the
plot of the free energy (in units of $\hbar\, \omega_C$) as a
function of the reduced temperature $t=k_B T/(\hbar\, \omega_C)$
in Fig.1. The curve was computed by taking $ m=0.8$ and
$\omega_R(t)=5 \,t^2\,\omega_C$. As we see, there is a region of
temperatures $t$ where the slope of the curve is positive, which
corresponds to a negative entropy. This is not necessarily a
problem, though, because what needs to be positive is the {\it
total} entropy of the system, which includes the self-entropies of
the wires. Each wire, being now an $RLC$ oscillator, has a free
energy ${\cal F}_{\rm self}$ equal to: \be {\cal F}_{\rm self}=k_B
T\,\log[1-\exp(-\hbar \omega_C/(k_B T))]\;.\ee As we have checked
numerically, inclusion of the wires self-entropies   makes the
total entropy of the system positive at all temperatures (we could
not obtain an analytical proof of this). The important conclusion to be drawn
is that inclusion of the antennas capacitances, however large they are,
is sufficient to resolve all thermodynamical inconsistencies discovered in the previous Section.

\section{Conclusions}

In recent years much attention has been devoted  to the study of
the thermal Casimir force between lossy metallic plates. Interest
in this problem arose as it was realized that even an
infinitesimal amount of dissipation is sufficient to produce a
large repulsive correction to the thermal Casimir force. In this
paper, we advocate the view that the abrupt change in the thermal
Casimir force   indicates that dissipation brings into the problem
a new physical phenomenon, characteristic of lossy conductors. We
argued that this phenomenon is indeed the well known
Johnson-Nyquist noise that exists in all conductors at finite
temperature. From this point of view, the repulsive correction to
the Casimir force can be simply explained as  the electrodynamic
interaction between the Johnson-Nyquist currents in either plate
and the eddy currents they induce in the other plate. To verify
this physical picture,   we  studied the electrodynamic forces
arising between two linear noisy antennas. As a first guess, we
solved the problem neglecting the capacitance of the antennas
associated with their end-points, under the reasonable assumption
that for separations between the antennas much smaller than the
antennas lengths, edge effects would be negligible. This kind of
assumption is customary in  treatments of the Casimir effect,
where standard Lifshitz theory assumes plates with infinite
extension.  After doing this approximation, we found a {\it
repulsive} force that shares all the features  found in the
Casimir problem,   in that it persists in the zero-resistance
limit, and it violates the Nernst heat theorem for antennas
without defects. These results appear to us as a clear indication
of the fact that the Johnson-Nyquist noise is the physical
phenomenon behind the large thermal correction to the Casimir
effect for lossy plates.

In the antennas model we found that inclusion of the antennas
capacitances, however large, drastically modifies the result, as
they make  the force vanish in the zero resistance limit and
remove all thermodynamic inconsistencies as well. Whether
capacitive edge effects are capable of resolving the analogous
difficulties met in the thermal Casimir effect requires further
investigations. In this connection, we remark that recently some
authors have questioned the compatibility of the existence of real
Johnson-Nyquist currents with Lifshitz theory, for the case of
finite plates of any size \cite{Geyer}. In our view, capacitive
edge-effects should be important only if the Johnson-Nyquist
current-fluctuations have a spatial size comparable to  the plates
size $L$. Detailed investigations of the thermal corrections to
the Casimir force \cite{lamor, esquivel} show that, at room
temperature, the important current fluctuations are transverse and
have characteristic frequency of order $\tilde{\omega}=\gamma \,
(\omega_c/\Omega_p)^2\,$, and characteristic spatial size of the
order of the plate separation $a$. Therefore,  we expect that
capacitive finite-size effects will be important, for plate
separations $a$ not too much  smaller than the plates size $L$.
While this is not the case in standard Casimir experiments, this
is indeed the typical situation in micro-mechanical devices, that
are currently under intense investigation (see the third of Refs.
\cite{bordag}).  At low temperature, the situation is more
hopeful. As the temperature is lowered in the cryogenic range, the
increasing degree of spatial correlation between the Johnson
currents, implied by the anomalous skin effect \cite{sveto}, leads
to a gradual suppression of fluctuations at small scales. It is
conceivable that at very low temperatures, independently of the
separation $a$, the current fluctuations become so  correlated as
to have a spatial extent comparable with the size of the plates.
When this point is reached, edge effects enter into play and may
become essential for a correct description, as discussed in this
paper.


\section*{References}

\begin{thebibliography}{99}

\bibitem{bordag}  Bordag M, Mohideen U and  Mostepanenko V M 2001
{\it Phys. Rep.} {\bf 353} 1

\item[]Milton K A 2004 {\it J. Phys.} A {\bf 37} R209

\item[]   Parsegian V A 2005 {\it Van der Waals Forces} (Cambridge
University Press, Cambridge, England)



\item[] Capasso F, Munday J N,  Iannuzzi D and Chan H B 2007 {\it
IEEE J. Sel. Top. Quant. Electron.} {\bf 13} 400

\bibitem{bimonte} Bimonte G 2007 {\it New J. Phys.} {\bf 9} 281



\bibitem{obrec} Obrecht J M, Wild R J, Antezza M, Pitaevskii L P,
Stringari S and Cornell E Q 2007 {\it Phys. Rev. Lett.} {\bf 98}
063201



\bibitem{lifs} Lifshitz E M 1956 {\it Sov. Phys. JETP} {\bf 2} 73
\item[] Lifshitz E M and Pitaevskii L P 1980 {\it Landau and
Lifshitz Course of Theoretical Physics: Statistical Physics} Part
II (Butterworth-Heinemann)

\bibitem{sernelius} Bostrom M and Sernelius B E 2000
{\it Phys. Rev. Lett.} {\bf 84} 4757

\bibitem{romero} Bezerra V B, Klimchitskaya G L and
Mostepanenko V M 2002 {\it Phys. Rev.} A {\bf 65} 052113
\item[]Bezerra V B, Klimchitskaya G L, Mostepanenko V M and Romero
C 2004 {\it Phys. Rev.} A {\bf 69} 022119

\bibitem{most3} Geyer B, Klimchitskaya G L and Mostepanenko V M
2005 {\it Phys. Rev.} D {\bf 72} 085009

\bibitem{most2} Klimchitskaya G L and
Mostepanenko V M 2006 {\it Contemp. Phys.} {\bf 47} 131

\bibitem{milton}  Brevik I,  Ellingsen L A and
Milton K A 2006 {\it New J. Phys.} {\bf 8}  236

\bibitem{decca}   Decca R S et al. 2005 {\it Ann. Phys.} {\bf 318}
37 \item[] Decca R S et al 2007 {\it Phys. Rev.} D {\bf 75} 077101
\item[] Decca R S et al.
2007 {\it Eur. Phys. J.} {\bf C 51} 963

\bibitem{lamor} Torgerson J R and Lamoreaux S K 2004  {\it Phys. Rev.} E {\bf  70}
 047102 \item[]Lamoreaux S K 2005 {\it Rep. Prog. Phys.} {\bf 68} 201
\item[]Bimonte G 2006 {\it Phys. Rev.} E {\bf 73} 048101

\bibitem{john} Johnson J B 1928 {\it Phys. Rev.} {\bf 32} 97

\item[] Nyquist H 1928 {\it  Phys. Rev.} {\bf 32} 110

\bibitem{jacks} Jackson J D 1999 {\it Classical Electrodynamics} (New York: Wiley)


\bibitem{kittel} Kittel C 1996 {\it Introduction to Solid State
Physics} (New York: Wiley).

\bibitem{Geyer} Geyer B, Klimchitskaya G L and Mostepanenko V M
2007 {\it J. Phys. A: Math. Theor.} {\bf 40} 13485

\bibitem{esquivel} Svetovoy V B and Esquivel R 2006 {\it J. Phys. A: Math. Gen.} {\bf
39} 6777















\bibitem{sveto} Svetovoy V B and Esquivel R 2005 {\it Phys. Rev.} E {\bf 72} 036113



\end{thebibliography}
\end{document}